\documentclass[manuscript]{emulateapj}
\usepackage{graphics}
\usepackage{graphicx}
\usepackage{rotating}
\usepackage{mathtools}
\usepackage{footmisc}
\begin{document}

\newcommand{\kms}{\mbox{km~s$^{-1}$}}
\newcommand{\s}{\mbox{$''$}}
\newcommand{\mloss}{\mbox{$\dot{M}$}}
\newcommand{\mdot}{\mbox{$\dot{M}$}}
\newcommand{\mdotaccr}{\mbox{$\dot{M}_a$}}
\newcommand{\my}{\mbox{$M_{\odot}$~yr$^{-1}$}}
\newcommand{\ls}{\mbox{$L_{\odot}$}}
\newcommand{\um}{\mbox{$\mu$m}}
\newcommand{\ujy}{\mbox{$\mu$Jy}}
\newcommand{\ms}{\mbox{$M_{\odot}$}}

\newcommand{\vexp}{\mbox{$V_{\rm exp}$}}
\newcommand{\vsys}{\mbox{$V_{\rm sys}$}}
\newcommand{\vlsr}{\mbox{$V_{\rm LSR}$}}
\newcommand{\tex}{\mbox{$T_{\rm ex}$}}
\newcommand{\teff}{\mbox{$T_{\rm eff}$}}
\newcommand{\tmb}{\mbox{$T_{\rm mb}$}}
\newcommand{\trot}{\mbox{$T_{\rm rot}$}}
\newcommand{\tkin}{\mbox{$T_{\rm kin}$}}
\newcommand{\dens}{\mbox{$n_{\rm H_2}$}}
\newcommand{\bri}{\mbox{erg\,s$^{-1}$\,cm$^{-2}$\,\AA$^{-1}$\,arcsec$^{-2}$}}
\newcommand{\brib}{\mbox{erg\,s$^{-1}$\,cm$^{-2}$\,arcsec$^{-2}$}}
\newcommand{\flux}{\mbox{erg\,s$^{-1}$\,cm$^{-2}$\,\AA$^{-1}$}}
\newcommand{\gsim}{\raisebox{-.4ex}{$\stackrel{>}{\scriptstyle \sim}$}}
\newcommand{\lsim}{\raisebox{-.4ex}{$\stackrel{<}{\scriptstyle \sim}$}}
\newcommand{\ha}{\mbox{H$\alpha$}}
\newcommand{\codos}{$^{12}$CO}
\newcommand{\cotres}{$^{13}$CO}
\newcommand{\hcntres}{H$^{13}$CN\,J=4--3}
\newcommand{\hcnmol}{H$^{13}$CN}
\newcommand{\paani}{H$_2$O}
\newcommand{\soocho}{SO\,N,J=8,8-7,7}
\newcommand{\csocho}{CS\,J=7-6}
\newcommand{\siodonau}{$^{29}$SiO\,J=8-7\,(v=0)}
\newcommand{\sodos}{SO$_2$\,4(3,1)-3(2,2)}
\newcommand{\soteenchar}{$^{34}$SO\,N,J=8,7-7,6}
\newcommand{\pscalunit}{g\,cm\,s$^{-1}$}
\newcommand{\cdensunit}{cm$^{-2}$}
\newcommand{\vdensunit}{cm$^{-3}$}
\newcommand{\hmol}{H$_2$}

\title{ALMA Observations of the Water Fountain Pre-Planetary Nebula IRAS\,16342-3814: High-velocity bipolar jets and an Expanding Torus}
%\author{R. Sahai\altaffilmark{1} et al.}
\author{R. Sahai\altaffilmark{1}, W.H.T. Vlemmings\altaffilmark{2}, T. Gledhill\altaffilmark{3}, C. S{\'a}nchez Contreras\altaffilmark{4}, 
E. Lagadec\altaffilmark{5}, L-\AA~Nyman\altaffilmark{6}, G. Quintana-Lacaci\altaffilmark{7}}

\altaffiltext{1}{Jet Propulsion Laboratory, MS\,183-900, California Institute of Technology, Pasadena, CA 91109, USA}
\altaffiltext{2}{Department of Earth and Space Sciences, Chalmers University of Technology, Onsala Space Observatory, SE-43992 Onsala, Sweden}
\altaffiltext{3}{Centre for Astrohysics Research, University of Hertfordshire, College Lane, Hatfield AL10 9AB, UK}
\altaffiltext{4}{Astrobiology Center (CSIC-INTA), ESAC campus, E-28691 Villanueva de la Canada, Madrid, Spain}
\altaffiltext{5}{Universit{\'e} C{\^o}te d'Azur, Observatoire de la C{\^o}te d'Azur, CNRS, Lagrange, France}
\altaffiltext{6}{Joint ALMA Observatory (JAO), Alonso de Cordova 3107, Vitacura, Santiago de Chile, and
European Southern Observatory, Alonso de Cordova 3107, Vitacura, Santiago, Chile}
\altaffiltext{7}{Instituto de Ciencia de Materiales de Madrid, Sor Juana Ines de la Cruz, 3, Cantoblanco, 28049, Madrid, Spain}

\email{raghvendra.sahai@jpl.nasa.gov}
\begin{abstract}
We have mapped \codos~J=3--2 and other molecular lines from the ``water-fountain" bipolar pre-planetary 
nebula (PPN) IRAS\,16342-3814 with $\sim0\farcs35$ resolution using ALMA. 
We find (i) two very high-speed knotty, jet-like molecular outflows, (ii) a central high-density ($>few\times10^6$\,\vdensunit), expanding torus of diameter 
1300\,AU, and (iii) the circumstellar envelope 
of the progenitor AGB, generated by a sudden, very large increase in the mass-loss rate to   
$>3.5\times10^{-4}$\,\my~in the past $\sim$455\,yr. Strong continuum emission at 0.89\,mm from a central source (690\,mJy), if due to thermally-emitting dust, 
implies a substantial mass (0.017\,\ms) of 
very large ($\sim$mm-sized) grains. The measured expansion ages of the above structural components imply   
that the torus (age$\sim$160\,yr) and the younger high-velocity outflow (age$\sim$110\,yr) were formed soon after the sharp increase in the AGB mass-loss rate. 
Assuming a binary model for the jets in IRAS\,16342, the high momentum rate for the dominant jet-outflow in IRAS\,16342 implies a 
high minimum accretion rate, ruling out standard 
Bondi-Hoyle-Lyttleton wind accretion and wind Roche lobe overflow (RLOF) models with white-dwarf or main-sequence companions. 
Most likely, enhanced RLOF from the primary or accretion modes operating within common envelope evolution are needed.
%The temporal gap between the EHV and HVO of $\sim100$\,yr, suggests that bipolar jet activity that likely 
%began somewhat before torus formation, was interrupted during the latter.

%TEN NEW WORDS ARE ADDED HERE TO TEST WORD COUNT

%ALMA mapping of the CO line emission in IRAS16342, which will enable us to determine the temperature and mass of accelerated
% material as a function of radius, latitudinal angle and velocity, is now needed to understand
%the jet-sculpting process and build quantitative models of jet-shaping.
\end{abstract}


\section{Introduction}\label{intro}
High-resolution imaging surveys of planetary nebulae (PNe) and pre-planetary nebulae (PPNe) reveal a dazzling variety of
bipolar and multipolar morphologies in these objects (e.g., Sahai, Morris \& Villar 2011; Sahai et al. 2007, Si{\'o}dmiak et al. 2008). 
In contrast, their progenitor AGB stars are
typically embedded in roughly spherical gas-dust circumstellar envelopes (CSEs) resulting from dense, slow (typically $\sim$\,10-25\,\kms) stellar
winds ejected with rates up to 10$^{-4}$\,\my~(e.g., Neri et al$.$ 1998, Castro-Carrizo et al. 2010). Amongst various mechanisms which have been considered (Balick \& Frank 2002),
hydrodynamic sculpting of a prior, spherically-symmetric outflow by wandering and/or episodic jets from the inside-out (Sahai \&
Trauger 1998) is widely believed to be the primary mechanism that can produce the extreme aspherical shapes in PNe and PPNe. 
Support for this mechanism comes from an increasing number of detailed observational studies of specific PPNe 
(Cox et al. 2000, Sahai et al. 2006,\,2013, Olofsson et al. 2015) and numerical
simulations of jet-CSE interactions (e.g., Lee \& Sahai 2003, Balick et al. 2013).

The key to understanding the jet-sculpting (or alternative) shaping mechanisms lies in the study of the structure and kinematics of the youngest 
objects: the ``water-fountains"  -- a class of
young PPNe in which unusually fast radial H$_2$O outflows with $V_{exp}\gsim$55\,\kms (Likkel \& Morris 1988, Likkel et
al. 1992) show that the jet activity is extremely recent ($\lsim\,100$ yr: Imai 2007). 
%These offer the clearest view of the ejection geometry, and in 
%the earliest phases the interactions of the high speed ejecta with the surroundings are at a maximum.

IRAS\,16342-3814 (hereafter IRAS\,16342) is the best studied and nearest ($\sim$2\,kpc) water-fountain PPN. Its morphology is well-resolved with optical (HST) and 
near-infrared (Keck Adaptive Optics) imaging. Radio interferometry (VLA, VLBA) shows water masers spread over a range of radial velocities encompassing 270\,\kms 
(Sahai et al. 1999 [Setal99], Claussen, Sahai \& Morris 2009). Single-dish mm-wave CO J=2--1 and 3--2 observations (He et al. 2008, Imai et al. 2012) reveal a 
massive, high-velocity molecular outflow. 
%The inclination of the jet axis to the line-of-sight, 44\arcdeg, is reasonably well-determined (Claussen, Sahai \& Morris 2009). 
The radio and optical data on IRAS\,16342 suggest nebular shaping in 
action, with a fast collimated outflow
creating a pair of bubble-like cavities by plowing into a surrounding dense, more slowly expanding AGB CSE.
% the lobes are these cavities which show up as reflection nebulae illuminated by starlight escaping through polar holes in a dense, dusty waist obscuring the central star. 
The near-IR AO imaging reveals a remarkable corkscrew structure inscribed on the lobe walls -- inferred to
be a signature of a precessing jet (Sahai et al. 2005: Setal05). Shock-excited \hmol~emission is found in the lobes, presumably excited 
by the jet (Gledhill \& Forde 2012). The central star remains obscured behind a dark equatorial waist even in the mid-IR (12\,\micron) (Verhoelst et al. 2009).
%VLA maps of the OH maser emission show features with the largest red- and
%blue-shifted velocities concentrated around the bright eastern and western polar lobes, respectively, whereas
%intermediate-velocity features generally occur at low latitudes, in the dark waist region (Sahai et al. 1999: S99). 

Here, we report millimeter-wave molecular-line and continuum interferometric 
observations of IRAS\,16342, using the Atacama
Large Millimeter/sub-millimeter Array (ALMA). These data clearly 
reveal, for the first time, the dynamics of a pair of young, high-speed, bipolar collimated outflows, the presence of a high-density central torus,  
and a surrounding circumstellar envelope resulting from an enormous increase in the mass-loss rate in the past few hundred years.
%{\it But in order to understand the dynamics and test theoretical models, high-resolution CO observations 
%are needed to determine the mass of gas associated with these features}.
%and mid-infrared (VISIR/VLT) (Lagadec et al. 2011) 
% He et al CO 2-1, 13CO 2-1; Imai CO 3--2, 13CO 3--2 (model has a r=1.5" halo, and a bipolar cavity+shell inside it)

\section{Observations}\label{obsreduce}
The data were obtained on 2015, June 13 with 37 antennas of the ALMA 12 m array in band 7. Using 35~min on the science target, the data cover four spectral windows 
(SPW's). Two 1.875~GHz-bandwidth SPW's were centered on 345.170~GHz (SPW0) and 331.214~GHz (SPW2) with 3840 channels each. Two additional continuum SPW's with 
2~GHz bandwidth with 128 channels were centered on 343.337~GHz (SPW1) and 333.104~GHz (SPW4). The baselines range from 21 to 783\,m implying a maximum recoverable 
scale of $\sim4$\arcsec. The data were calibrated using the ALMA pipeline using Ceres as flux calibrator and the quasars J1924-2914 ($\sim2.7$~Jy) and J1636-4102 
($\sim0.14$~Jy) as bandpass and gain calibrator respectively. Following the standard calibration, self-calibration was performed on the continuum emission of 
IRAS\,16342. Images were created using a Briggs weighting scheme channel averaging to $4$~km~s$^{-1}$ in continuum subtracted line SPW's. In the continuum SPW's the 
effective channel width corresponds to $\sim14$~km~s$^{-1}$. Typical channel rms is $\sim 2.3$~mJy~beam$^{-1}$ in line-free $4$~km~s$^{-1}$ channels.
%Typical channel rms is $\sim 2.3$~mJy~beam$^{-1}$ and $\sim0.8$~mJy~beam$^{-1}$ in line-free $4$ and $14$~km~s$^{-1}$ channels.

\section{Results}\label{obsresult}
We have identified a number of lines in the 4 SPW's. In addition to the J=3--2 lines of \codos~(SPW0) and \cotres~(SPW2), and \hcntres~(SPW0) that are discussed here, 
we also find \soocho, (SPW0, SPW1), \csocho~(SPW1), \siodonau~(SPW1), \sodos~(SPW3), and \soteenchar~(SPW3) (discussed in Sahai et al. (2017, {\it in prep.})).
\subsection{High-Velocity Outflows}
\begin{figure}[htb]
%\resizebox{0.45\textwidth}{!}{\includegraphics{13co32-cont-band7-pap.pdf}
%\resizebox{0.55\textwidth}{!}{\includegraphics{i16342-co32-cen-hiredblu.pdf}
\includegraphics[scale=0.5]{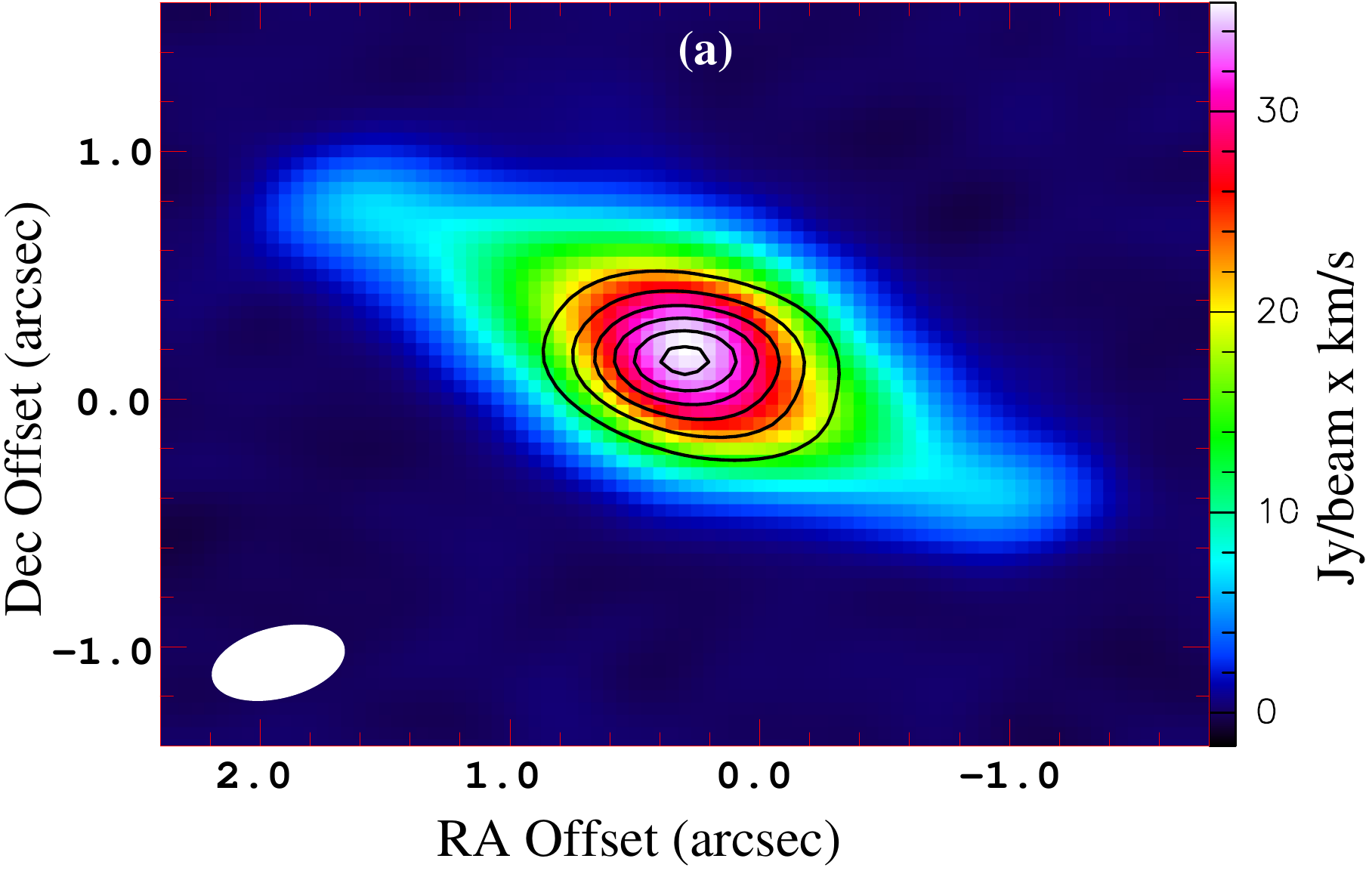}
\includegraphics[scale=0.5]{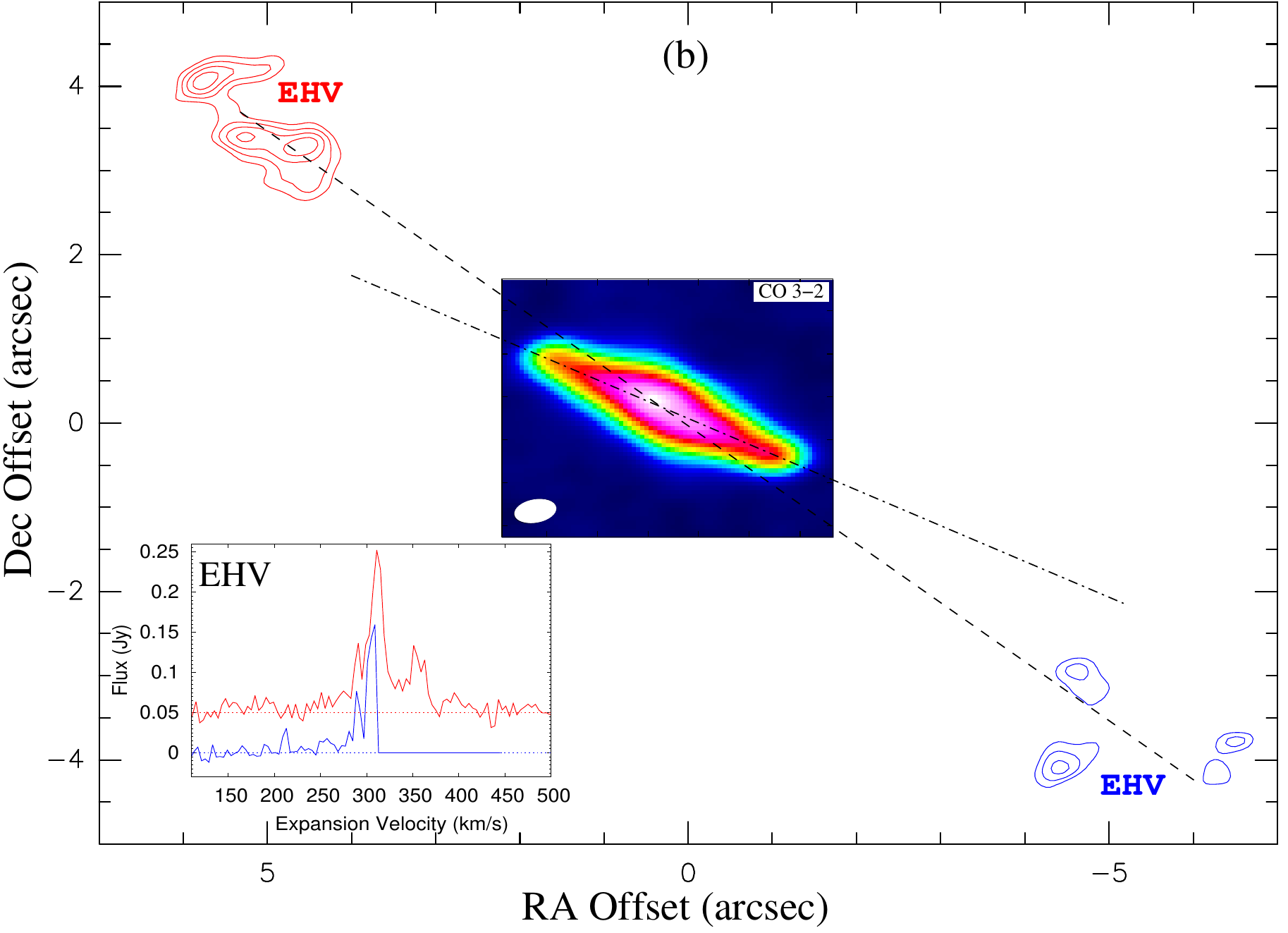}
%\caption{Moment 0 maps of the \codos~and \cotres~emission from IRAS\,16342.}
\caption{ALMA maps of the \cotres, \codos~J=3--2 and 0.89\,mm continuum emission from IRAS\,16342 (white ellipses show the beam FWHM and orient). 
(a) \cotres~emission integrated over the HVO velocity 
range of $\pm$260\,\kms~centered at the systemic velocity. The continuum intensity is shown as contours, with lowest level (step) 
equal to 20\% (15\%) of the peak, 0.25\,Jy\,beam$^{-1}$. The continuum peak is located 
at (J2000) RA=16:37:39.935, Dec=-38:20:17.15, and the phase center (i.e., offset 0,0) at RA=16:37:39.91, Dec=-38:20:17.3. (b) Colorscale shows \codos~emission 
integrated 
over the HVO velocity range. Contours show red and blue-shifted 
components of the EHVO covering emission at $\pm$310\,\kms~relative to the systemic velocity (the red emission is integrated over 34\,\kms, whereas the 
blue emission lies at the edge of the bandpass, is integrated over 24\,\kms). Inset shows the \codos~spectra 
of the red- and blue-shifted components of the EHVO, and dashed (dash-dotted) line shows the axis of the EHVO (HVO).}
%The beam size (FWHM) for the \codos~(\cotres) J=3--2 data is $0\farcs49\times0\farcs27$, $PA=-78.1$\arcdeg, ($0\farcs54\times0\farcs28$, $PA=-76.3$\arcdeg)}
\label{co32mom0}
\end{figure}

The IRAS\,16342 \codos~and \cotres~moment-zero maps, generated by integrating over a velocity range $\pm260$\,\kms~relative 
to the systemic velocity ($V_{lsr}\sim45$\,\kms), show point-symmetric shapes (Fig.\,\ref{co32mom0}a). The SW (NE) lobe is blue- (red-) shifted, and 
the nebula's long axis is 
oriented at $PA=67$\arcdeg. The \codos~image shows emission from an extreme high-velocity outflow (EHVO) located much further from the nebular tips at a (projected) 
radial velocity of $\sim\pm$310\,\kms~(Fig.\,\ref{co32mom0}b) and the axis of this 
outflow is oriented at $PA\sim55$\arcdeg. Since this emission has a high degree of symmetry relative to the center, 
in both position and velocity, and because it lies along an axis close to the nebular axis, we conclude that it 
is related to IRAS\,16342, and not to random interstellar clouds in the Galactic plane. 
%This emission represents an earlier ejection. separation about 12", time=6"/sin 43 / V/cos 43 = 1.8e17/310e5 tan 43 = 197 yr.
% CSE expansion time-scale at $R_{out}$ is $t_{out}=240$\,yr(45/Vagb)
% EHV, separation about 12", time=6"/sin 43 / V/cos 43 = 1.8e17/ (310e5 tan 43) = 197 yr.
% torus 0.325 arcsec*3e16/(19e5 tan 43) = 174 yr
% HV at lobe tip 110 yr
A comparison of the expansion velocity ranges of the blue- and red-shifted EHVO components (Fig.\,\ref{co32mom0}b, inset), 
indicates that part of the former lies beyond the bandpass edge.

\begin{figure}[htb]
%\resizebox{0.95\textwidth}{!}{\includegraphics{i16342-12co-h13cn-pv-blobs_flipx_spec5}
\includegraphics[scale=1.0]{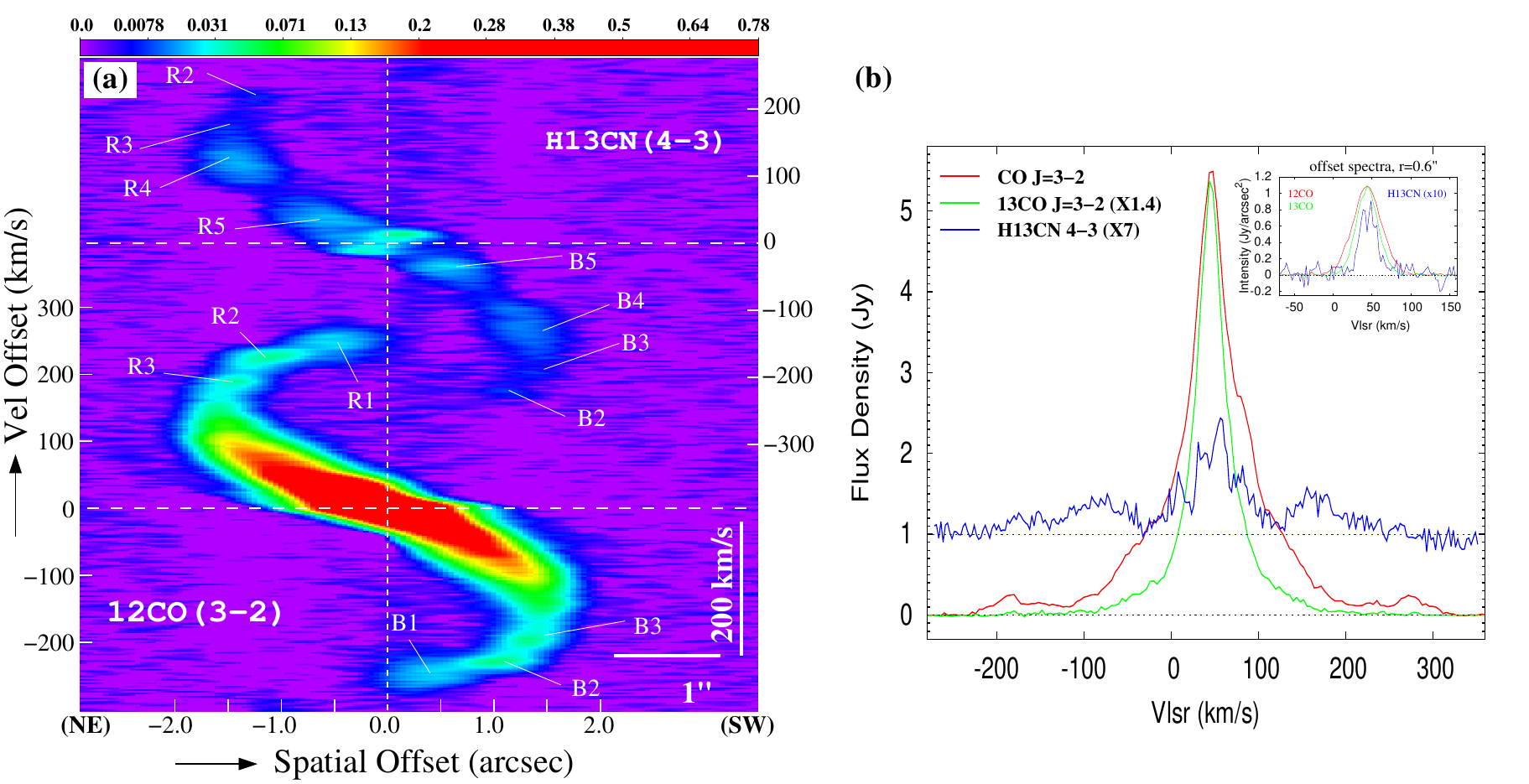}
\caption{(a) Position-velocity plot of the \codos~J=3--2 and \hcntres~emission from IRAS\,16342 along $PA=67\arcdeg$. X-axis shows spatial offset from 
the nebular center (marked by dashed vertical line). Y-axis shows velocity offset from the systemic velocity ($V_{lsr}=45$\,\kms, marked 
by dashed horizontal lines). (b) Spectra of the spatially-integrated 
(total) flux density for \codos J=3--2, \cotres J=3--2 (scaled by factor 1.4) and \hcntres~(scaled by factor 7 and shifted upwards by 1 Jy). 
The contribution of emission from the red-wing \codos~J=3--2 to that of the blue-wing of \hcntres~(and vice cersa) has been subtracted. Inset shows 
the mean \codos~and \cotres~J=3--2 spectra extracted from $0\farcs15$ diameter apertures located at $r=\pm0\farcs6$, 
and the mean \hcntres~spectrum extracted from annular wedges with radius (width) 
$\pm0\farcs6$ ($0\farcs12$) and opening angle of $52\arcdeg$, along $PA=133\arcdeg$.}
\label{coh13cnpv}
\end{figure}

Position-velocity (PV) plots of \codos~J=3--2 and \hcntres~emission along the nebular axis 
generally show an S-shape, with small knots (Fig.\,\ref{coh13cnpv}a). We have identified a total of five pairs of such knots ($B1-B5$ and $R1-R5$). 
The emission from the central bipolar nebula extends 
over $-230<V_{lsr}(\kms)<325$ in \codos~(Fig.\,\ref{coh13cnpv}b); the emission from \cotres~J=3--2 and \hcntres~is weaker, and 
is detectable over a somewhat smaller range ($\sim500$\,\kms).
% A fourth is seen at a lower offset velocity about $V_{lsr}=-90$\,\kms on the blue side in the \cotres~plot. 
% central emission in CO looks most symmetric at V=45.65 km/s in SPW0.fits

\begin{figure}[htb]
%\resizebox{0.5\textwidth}{!}{\includegraphics{/u4/sah1/sahai/optdata/wfpc2/iras16342/blob-geom-pap}
%\resizebox{0.5\textwidth}{!}{\includegraphics{blob-geom-pap}
\includegraphics[scale=0.5]{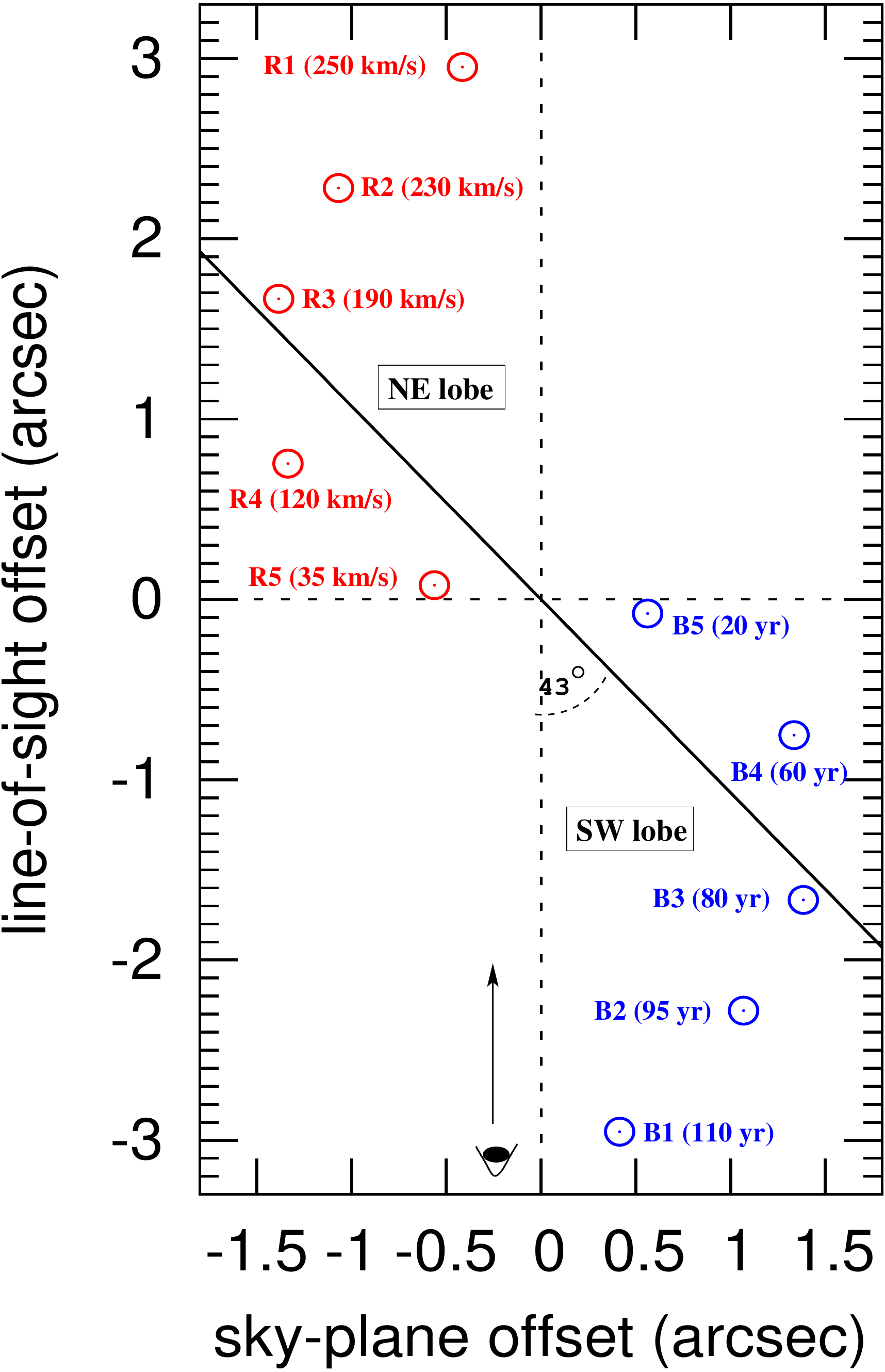}
\caption{Blob locations in the SW and NE lobes along the line-of-sight ($los$), derived assuming a uniform outflow velocity ($250$\,\kms). Blob ages (radial 
velocities), which are identical for identically numbered blobs, are shown for the SW (NE) lobe. Since we have assumed that the outflow velocity is the same as 
that of blobs $B1$,\,$R1$, we cannot directly derive their $los$ offsets; their adopted locations are based on other considerations (see text). 
The sloping black line shows the axis of the central torus.}
\label{blobgeom}
\end{figure}

%that are closed at their tips, as seen in optical and near-IR images. 
The spiral structure seen in the PV plots most likely indicates emission from a precessing high-velocity bipolar outflow (HVO) 
that entrains material in the near and far bipolar lobe walls.
%at progressively larger radial offsets
Blobs $B1$ and $R1$ have the highest outflow velocities and the smallest spatial offset from 
the center, and are thus moving along an axis close to the line-of-sight ($los$), implying an outflow speed of slightly more than 250\,\kms. If we assume 
that all blobs are moving radially with this same outflow speed, i.e., 250\,\kms, then we can derive the locations of blob pairs $B2-R2$, 
$B3-R3$, $B4-R4$, and $B5-R5$, along the $los$ and their ages ($\sim$95,\,80,\,60,\,20\,yr, respectively: Fig.\,\ref{blobgeom}). 
The $B1-R1$ pair's much lower brightness, compared to that of the other pairs, suggests that $B1-R1$ is the oldest pair, since we expect the 
blobs to cool and dissipate with time. Assuming that the interval between the ejections of $B1-R1$ and $B2-R2$ is similar to 
that between $B2-R2$ and $B3-R3$, the $B1-R1$ pair are about 110\,yr old. The 
corresponding inclination angle\footnote{all inclination angles, here and elsewhere in the paper, are relative to the $los$} 
for the $B1-R1$ axis is then $8\arcdeg$, and the resulting locations of $B1$ and $R1$ are consistent with 
an extrapolation of the S-shape defined by the 
other blobs. We conclude that the HVO expansion age is 
$\sim110$\,yr. Note that since the kinematics of the CO emitting blobs may depend on the 
hydrodynamic interaction between the underlying jet and surrounding CSE 
material, the assumption of a uniform radial outflow velocity is likely not strictly true, and the derived ages are somewhat uncertain. 

%$\theta_{(B1-R1)}=
%       ProjVel     theta  age
%B2-R2  227         23.7
%B3-R3  193         38.9
%R4,B4  120         61
Excluding the velocity range $0<V_{lsr}(\kms)<90$ which 
includes emission from the AGB CSE (see \S\,\ref{agbcse}), we find that the velocity-integrated fluxes in the HVO's red-wing 
($90<V_{lsr}(\kms)<320$) and blue-wing ($-228<V_{lsr}(\kms)<0$)   
are 103.9 and 89.8 Jy-\kms, respectively.
%The tips of the lobes are located at a radial offset of $1\farcs5$.
%The dense walls of the lobes are optically thick in \codos~and we only see the near-surfaces of each wall.

The \cotres~to \codos~line intensity ratio is almost unity towards the center, implying a very low relative 
abundance ratio, $f_{12/13}$=f(\codos)/f(\cotres). 
Hence, here and in the following sections, we assume 
that $f_{12/13}\sim3$, the lowest value achievable via equilibrium CNO nucleosynthesis, and take 
f(\codos)=$3\times10^{-4}$, or f(\cotres)=$10^{-4}$. Larger values of $f_{12/13}$ will imply proportionately larger values 
of mass and mass-dependent quantities.
%(scalar momentum, energy, mass-loss rates). 
%The lowest \cotres~to \codo ~intensity ratios measureable are about ($\sim$0.2) in the blue- and red- lobe tips, which suggest

%It decreases radially most rapidly along the nebular axis; whereas 
%along the equatorial plane, it remains close to unity. This clearly supports the our inference that there is high-density, high optical-depth  
%material along the equatorial plane, as suggested by the presence of the \hcntres~torus.
% cos 40 = 0.766, cos 43 = 0.731
% $P_{HVO}>1.66\times10^{38}$ for theta=40 deg, $P_{HVO}>1.74\times10^{38} for theta=43
% $E_{HVO}>2.89\times10^{45}$ for theta=40 deg, $P_{HVO}>3.18\times10^{38} for theta=43
The scalar momentum and kinetic energy in the HVO, derived   
from the \codos~J=3--2 data, assuming the emission to be optically-thin (and distance 2\,kpc), is 
$P_{HVO}>1.7\times10^{38}$\,\pscalunit~and $E_{HVO}>3.2\times10^{45}$\,erg, using the formulation 
described in Bujarrabal et al. (2001). We adopt an inclination angle of $\theta=43\arcdeg$ (see \S\,\ref{torustext}). 
The HVO mass is, $M_{HVO}=6.2\times10^{-3}$\,\ms. 
We assume a uniform excitation temperature, $T_{exc}=15-30$\,K (resulting in a $\sim\pm20$\% uncertainty in 
$P_{HVO}$, $E_{HVO}$, and $M_{HVO}$). The HVO 
mass-loss rate is $>5.6\times10^{-5}$\,\my. For the EHVO we find 
% assuming its blue-component is similar to the red one,
that the mass, scalar momentum ($P_{EHVO}$), and kinetic energy ($E_{EHVO}$) are 7\%, 21\%, and 51\% of the corresponding values for the HVO; 
its deprojected speed is $360-540$\,\kms, and expansion age is $\sim130-305$\,yr, allowing for the EHVO's inclination angle to vary from that of the HVO by 
$\sim\pm12\arcdeg$, similar to the difference in their position angles. 

%$t_{HVO}=r_{R4,B4}/(V_{R4,B4}$\,tan\,$43\arcdeg)\sim110$\,yr, where $r_{R4,B4}=2\farcs8$ and $V_{R4,B4}=240\,\kms$ 
%  are the radial and velocity separations between blobs R4 and B4 that lie near the lobe tips

%EHV expansion age 6" x 3e16 / 310e5 tan 43 = 197.363 yr; spread is 6" x 3e16 / 310e5 tan (43+/-12) = 129-306 yr
%EHV inclination theta 31 to 55 deg; Vexp=310/cos(theta)=361.7 to 540.5

The values of $P_{HVO+EHVO}$ and $E_{HVO+EHVO}$ in IRAS\,16342 
lie within the ranges of these quantities for PPNs, $10^{37-40}$ g\,cm\,s$^{-1}$ and $10^{42-46}$\,erg (Bujarrabal et al. 2001). 
The HVO cannot be driven by radiation
pressure because its dynamical (expansion) time-scale is much smaller than that required by radiation
pressure to accelerate the observed bipolar outflow to its current speed, $t_{rad}>P_{HVO}/(L/c)\sim 7.3\times10^3$ yr, 
%$t_{rad}=P_{sc}/(L/c)\sim 3.3\times10^4$ yr (x2 because scalar mom x2 above)
given IRAS\,16342's luminosity of $L_{*}=6000$\,\ls.

\subsection{A Central Torus}\label{torustext}
\begin{figure}[htb]
%\resizebox{0.9\textwidth}{!}{\includegraphics{h13cn-torus}
\includegraphics[scale=1.0]{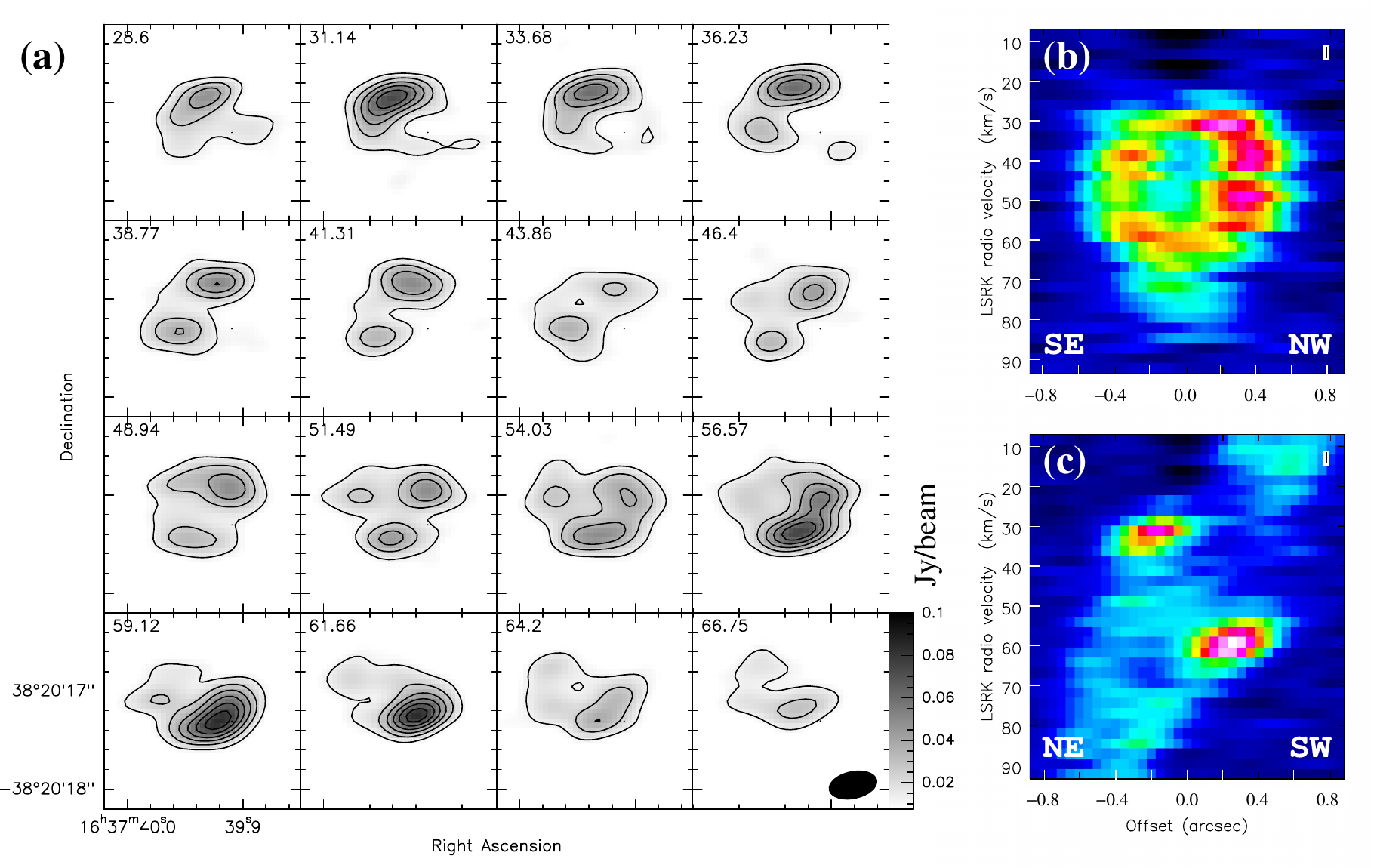}
\caption{\hcntres~emission from IRAS\,16342: (a) channel maps near the systemic velocity, showing a tilted torodial structure with 
major axis along $PA=133\arcdeg$ (black ellipse in $V_{lsr}=66.75$\,\kms~panel 
shows beam FWHM and orient, and the minimum contour level and step is $0.0125\times10^{-2}$\,Jy\,beam$^{-1}$); 
position-velocity cuts along the (b) major axis, and (c) minor axis of the torus.}
\label{h13cntorus}
\end{figure}

The \hcntres~emission in channels near the systemic velocity, $V_{lsr}=43.86-46.4$\,\kms~(Fig.\,\ref{h13cntorus}a) shows 
two blobs along an axis at $PA\sim133$\arcdeg~implying an inclined, equatorial torus or squat-cylindrical structure. 
%(i.e. confined to low-latitudes)
The panels at $V_{lsr}=64.2-61.66$\,\kms ($V_{lsr}=31.14-28.6$\,\kms)~presumably show the red-shifted (blue-shifted) parts of this structure to the SW (NE), 
consistent with it being an equatorial structure that engirdles the bipolar lobes.
%, which show the opposite pattern of red and blue-shift, with the SW (NE) lobe being blue (red) shifted. 
A PV plot along the torus major axis (Fig.\,\ref{h13cntorus}b) 
shows the characteristic elliptical shape expected for an expanding torus (since the beam-size has a relatively large extent 
at $PA\sim43$\arcdeg, the cut shows emission from all regions of the torus). The observed PV plot along the minor 
axis (Fig.\,\ref{h13cntorus}c) shows the 
approaching (receding) parts of the torus. Using the tilted torus's projected size along its 
major axis (0\farcs65) and minor axis (0\farcs44) as measured from the PV plots, 
we find that the torus size is 1300\,AU, and its inclination is 43\arcdeg, consistent with that of 
%inclined to the line-of-sight by $\theta _t=sin^{-1}(0.44/0.65)
the bipolar lobes, $\sim40\arcdeg$ (Setal99) and the high-velocity \paani~outflow axis, $\sim45\arcdeg$ (Claussen et al. 2009). The 
deprojected torus expansion velocity is, $V_{tor}=20$\,\kms, and its expansion time-scale is 160\,yr.
%the H13CN lines are going to be a good (relatively) optically-thin probe of high-density material in this object in the future 
%contrast between torus and neighbouring AGB CSE
%rms in h13cn is 0.0022 Jy/beam, peak is 0.048 Jy/beam (S/N=21.8)

We show that the torus density is quite high. The observed peak \hcntres~intensity in the torus 
at the systemic velocity is, 37.5\,mJy\,beam$^{-1}$ ($T_b\sim8.8$\,K), implying a minimum \hcnmol~column 
density of, N(\hcnmol)=$4\times10^{15}$\,\cdensunit($\delta V/1\,\kms)$ ($\delta V$ is the intrinsic line-width), 
for optically-thin emission and a volume density adequate for exciting the \hcntres~line. 
%(n(\hmol)$\sim10^6$\,\vdensunit)
This result is not sensitive to the assumed excitation temperature, 65\,K, derived from the \codos~and \cotres~J=3--2 line intensities at 
the torus radius, $r\sim0\farcs32$. Taking the 
radiatively-connected column contributing to the line-center emission in the expanding torus to be 
% for a medium expanding radially at $V_{tor}$, 
$\delta z= r\,\delta V/V_{tor}$, we find $\delta z= 5\times10^{14}$\,cm\,($\delta V$/1\,\kms) at $r=0\farcs32$. If 
we assume the fractional abundance of \hcnmol~to 
be one-third that of HCN, f(HCN), then  
n(\hmol)=3\,N(\hcnmol)/(f(HCN)\,$\delta z)=1.1\times10^8$\,\vdensunit, taking f(HCN)=$2.2\times10^{-7}$, from observations of the 
oxygen-rich AGB star IK\,Tau (Decin et al. 2010). If we assume theoretical values, f(HCN)=$9.06\times10^{-5}$ (Cherchneff 2006) 
or f(HCN)$=2.12\times10^{-6}$ (Duari et al. 1999), then n(\hmol)=$2.7\times10^5-1.1\times10^7$\,\vdensunit. 
Assuming that the ratio of the peak torus \hcntres~intensity relative to that in the AGB CSE, (say) at $r=0\farcs75$ ($>14$) is comparable to 
the corresponding density ratio, we infer 
that the torus density is $>3.3\times10^6$\,\vdensunit, and f(HCN)$<7.3\times10^{-6}$ (the 
AGB CSE density at $r=0\farcs75$ is $\sim2.4\times10^5$\,\vdensunit, see \S\,\ref{agbcse}).

\subsection{The Central Continuum Source}
We detect a compact, marginally-resolved continuum source at the center\footnote{peak located at (J2000) RA=16:37:39.935, Dec=-38:20:17.15}, with (FWHM, 
undeconvolved) size $0\farcs64\times0\farcs48$, $PA=80$\arcdeg~(beam 
is $0\farcs50\times0\farcs26$, $PA=-78.2$\arcdeg). The peak intensity and integrated flux are 0.25 Jy/beam and 0.69 Jy at $\nu=338.19$\,GHz. 

The 0.87\,mm continuum flux density observed with the 
12-m APEX telescope's $18\farcs6$ beam, $602\pm90$\,mJy (Ladjal et al. 2010), 
%with a 0.15\,mm wide filter 
is within errors, the same as our 
observed value (690\,mJy). A 1.3\,mm flux density of $277\pm13$\,mJy (Guertler et al. 1996) was measured with the SEST 
$23{''}$ beam. Both these flux densities are significantly in excess of the values ($\sim113$ \& 30\,mJy at 0.886 \& 1.3\,mm) derived from an 
extrapolation of the 2-D dust radiative-transfer model fit to IRAS\,16342's SED and near-IR polarised light imaging (Fig.\,1 of Murakawa \& Izumiura 2012), 
in which the dominant mass component is a 1\,\ms~torus of radius $700-1000$\,AU that includes grains with a radius up to 10\,\micron.
%assuming $\lambda\,F_{\lambda}=5\times10^{-14}$\,W\,m$^{-2}\,(\lambda/300\,\micron)^{-3.5}$ 
%(where the longest observed wavelength is $160$\,\micron).\
% cont lambda is 0.886 mm

Assuming that the excess 0.89 and 1.3\,mm flux densities (577 and 247\,mJy) are due to thermal dust 
emission\footnote{The contribution of free-free emission is negligible given the upper limit on radio 
emission by Sahai et al. (2011)} from the same region, we find that 
the dust emissivity power-law ($\chi_{\nu}\propto\nu^{\beta}$) index is, $\beta\sim0.46-0.27$ for a dust temperature of, 
$T_d\sim30-120$\,K (e.g., $T_d\sim50$\,K for the large grains in the torus of the PPN, IRAS\,22036+5306: Sahai et al. (2006)). 
The characteristic dust emission radius, assuming complete absorption of the starlight by 
intervening dust very close to the star and re-radiation at an effective temperature of $T_{*}=650$\,K 
(Setal05), is given by $r_d=[\frac{L_{*}T_{*}^\beta}{16\pi\sigma}]^{1/2}T_d^{-(2+\beta/2)}$ (Setal99) and 
lies in the range $r_d\sim(2.0-0.08)\times10^{17}$\,cm.
%and a conservatively broad range of the stellar temperature, $T_{*}=3000-8000$\,K, we get $r_d\sim(2.9-0.1)\times10^{17}$\,cm. 
Since the observed outer radius of the dust emitting source, derived from its major-axis (FWHM) of $0\farcs64$, 
is $\sim0.1\times10^{17}$\,cm, we infer that $T_d\sim120$\,K and $\beta\sim0.27$. The low value of $\beta$ at mm/sub-mm 
wavelengths implies that the grains are very large ($>$few mm); taking  
$\chi_{\nu}(0.85\,mm)=1.4$\,cm$^{2}$g$^{-1}$ (Sahai et al. 2011), the dust mass of this large-grain component is, $M_d=0.017$\,\ms. 
Testing alternatives to the large-grain model, such as spinning 
dust and magnetic nanoparticles (e.g., Draine \& Hensley 2012), will 
require continuum measurements at additional mm-submm wavelengths.
%, and will be explored in Sahai et al. (2017, in prep.).
% average FWHM of source = 0.64 arcsec = 1.93e16 cm
% calibrated_final_cont_image_p0.pbcor.fits, peak is 0.248 Jy/beam
% calibrated_final_cont_image.image.fits, peak is 0.253 Jy/beam, same FWHM as above

% the 12co32 brightness temp is 89.7 K at 44 km/s Vlsr, 13co32 is 74.5 K at center
% the h13cn brightness temp is ~2-3 K at 44 km/s Vlsr in torus
% 100K, 0.5e5, 1e19, 35 km/s FWHM
% in blob B3, 12co/13co=8.04K/1.45K=5.54 (V=-152 km/s Vlsr, 7 x 5 pix ellipse, cen at 208.7, 181.45 pix)
% in blob R3, 12co/13co=7.23K/1.34K=5.4  (V= 236 km/s Vlsr, 7 x 5 pix ellipse, cen at 157, 203 pix)

\subsection{The Progenitor AGB Circumstellar Envelope}\label{agbcse}
\begin{figure}[htb]
%\rotatebox{270}{\resizebox{0.4\textwidth}{!}{\includegraphics{12co-13co32-radial-ch111_c43_c223_w26_cut_pap}}
\includegraphics[angle=0,origin=c,scale=0.5]{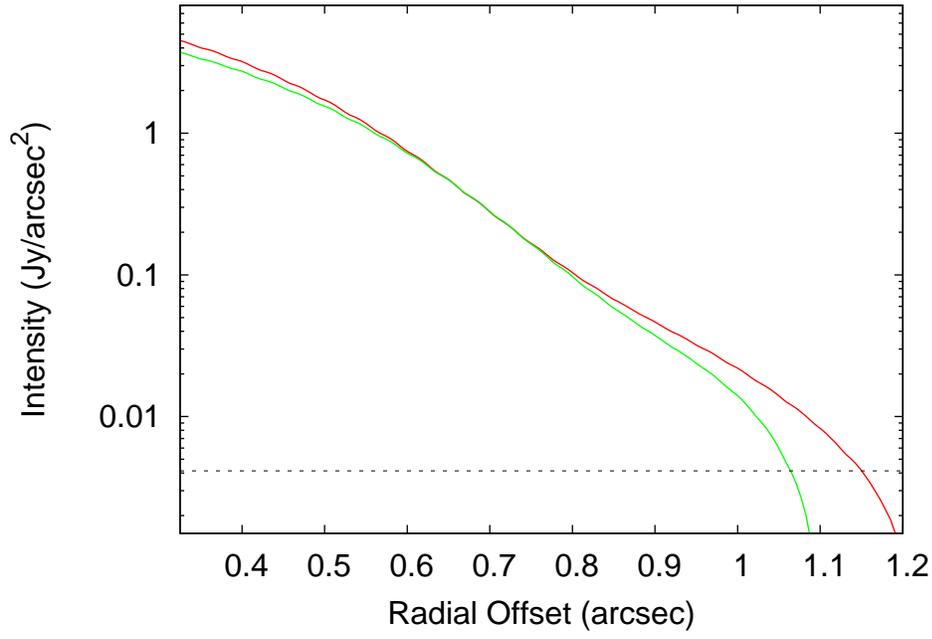}
\caption{The mean radial \codos~(red) and \cotres (green)~J=3--2 intensity in the central velocity channel ($V_{lsr}=44$\,\kms), 
in wedges of angular width $26\arcdeg$ oriented along $PA=133\arcdeg$ and $PA=-47\arcdeg$ (data from the two wedges have been averaged). The dashed horizontal line 
shows the 1$\sigma$ noise level.}
\label{coavint}
\end{figure}

Our ALMA data provide direct evidence for the jet, and the presence of a surrounding CSE, that are needed 
for creating the observed bipolar morphology in IRAS\,16342 as proposed earlier (Setal99, Setal05). In order to determine 
the CSE size, we first determine 
the radial direction along which the \codos~J=3--2 line profile is minimally affected by the bipolar outflow from an inspection of 
the spectra in regions offset from the center in a direction orthogonal to the bipolar nebula. We find that the 
profile remains centered at the systemic velocity at all positions along $PA\sim133\arcdeg$. 
We then determine the mean
\codos~and \cotres~intensities at $V_{lsr}=44$\,\kms, i.e., the spectral channel closest to the systemic velocity, 
in two wedges of angular width $26\arcdeg$ oriented along $PA\sim133\arcdeg$ and $PA\sim-47\arcdeg$ as a function of radius, and 
average these. The resulting intensity cut traces emission from the AGB CSE out 
to a radius, $R_{out}\sim1\farcs1$. At $r>R_{out}$, the intensity falls steeply (Fig.\,\ref{coavint}). 
% -- at a radial offset of 0\fracs6, the FWZI is $\sim90$\,\kms, implying a rather large expansion velocity, $V_{exp}=45$\,\kms

The mean \codos, \cotres~and \hcnmol~line profiles at $r=\pm0\farcs6$ 
along $PA\sim133\arcdeg$, have widths at their base (FWZI) of $\sim$100, 85, and 38\,\kms~(Fig.\,\ref{coh13cnpv}b, inset). 
It is likely that the optically-thick CO line profiles include contributions  
from low-density outflow material that does not emit significantly in the \hcntres~line. 
The narrow \hcntres~line emission can be seen out to $r\sim0\farcs75$; we conclude 
that the FWZI of the latter provides the best estimate of the AGB CSE 
expansion velocity, i.e., $V_{AGB}\sim19\,\kms$/[1-($0\farcs6/R_{out}$)$^2$]$^{0.5}\sim23$\,\kms.

The radial \cotres~and \codos~J=3--2 line intensities are very close to each other  
at all radii $r<\sim0\farcs85$, implying that both of these lines are quite optically thick in this region.
The increasing difference between the \cotres~and \codos~intensities for 
$r>\sim0\farcs85$ signifies a decrease in the optical depth of these lines. From
the observed  \codos~to \cotres~J=3--2 intensity ratio of 1.56 at $r=1{''}$, we derive 
a \cotres~J=3--2 optical depth of $\tau_{13}\sim1$, an excitation 
temperature $T_{exc}\sim4.1$\,K, and a mass-loss rate $\mdot_{AGB}>3.5\times10^{-4}\,(V_{AGB}/23\,\kms)$\,\my.

At $r>R_{out}$, both the \cotres~and \codos~J=3--2 intensities, and their ratio, 
$R_{I(13/12)}$, decrease rapidly -- e.g., for a 4\% increase in radius beyond $r=1\farcs0$, 
$R_{I(13/12)}$ decreases by 28\%. This decrease in 
$R_{I(13/12)}$ must result from a more rapid decrease in the \cotres~optical depth (and thus its density) compared 
to \codos, and cannot be due to interferometric loss of flux, especially since the maximum recoverable scale is $4{''}$.
Without a sharp decrease in density in this region ($\sim10^5$\,\vdensunit~at $r\sim R_{out}$),  $R_{I(13/12)}$ 
could not decrease rapidly because $T_{exc}$ for both \cotres~and \codos~J=3--2 would 
remain very close to $T_{kin}$, and a sharp decrease in $T_{kin}$ is unlikely.
We therefore conclude that there is preferential photodissociation of \cotres~over \codos~by the interstellar UV at $r\gtrsim1{''}$, implying that the 
mass-loss rate at $r\gtrsim R_{out}$ must be $< 0.7\times10^{-6}$\,\my\,($V_{AGB}$/23\,\kms)$^{0.5}$ (Mamon et al. 1988). The AGB expansion 
time-scale at $r=R_{out}$ is 455\,yr\,(23\,\kms/$V_{AGB}$).

\section{Discussion and Concluding Remarks}\label{discuss}
% CSE expansion time-scale at $R_{out}$ is $t_{out}=240$\,yr(45/Vagb)
% EHV, separation about 12", time=6"/sin 43 / V/cos 43 = 1.8e17/ (310e5 tan 43) = 197 yr.
% torus 0.325 arcsec*3e16/(19e5 tan 43) = 174 yr
% HV at lobe tip 110 yr
% assume torus cross-sectional radius = 0.1 * torus radius
% torus mass Area * 2 pi r * dens * 2 * 1.67e-24/1.989e33 = 9.86*(0.0325*0.325*0.325)*3e16**3 * 1e8 * 1.67e-24/1.989e33 = 0.077 Msun
Our ALMA study of IRAS\,16342 has characterised four different circumstellar components. The oldest of these is 
the AGB CSE ($\sim455$\,yr), followed by the EHVO ($130-305$\,yr) and torus ($160$\,yr) and the youngest is the HVO ($\sim110$\,yr). This sequence 
suggests that the torus and HVO were formed a few hundred years after a sharp increase in the mass-loss rate of the progenitor AGB star.
%The temporal gap between the EHV and HVO, suggests that bipolar jet activity, that likely began somewhat 
%before the torus formation, was interrupted during the latter. 
%The torus axis is significantly misaligned with the EHV and HVO axes, indicating that these are not hydrodynamically collimated by the former. 
The HVO (and possibly the EHVO) appears very soon after torus formation in IRAS\,16342. Huggins (2007) derives 
the same result for a small sample of similar objects (late AGB stars, 
PPNe and young PNe), and shows that it naturally favors the class of models in which a companion interacts 
with the central star. These include models where the build-up of a torus 
enhances the accretion rate in a disk around a companion that then drives jet-like outflows, or 
both spin-up and ejection of the stellar envelope of the primary occur during a common-envelope (CE) phase as the companion spirals in to the 
center of the AGB star. The detailed point-symmetry seen in IRAS\,16342's outflows suggests a stable precessing or wobbling disk, and 
argues against explosive scenarios (e.g., Matt et al. 2006).
% Huggins assuems formation of torus and disk are linked, so that torus formation assumed = disk formation age

If the jet activity in IRAS\,16342 is driven by an accretion disk around a companion, then we can constrain the type of accretor 
as done in the study by Blackman \& Lucchini (2014, BL14), who determine the minimum required accretion rates based on the 
properties of collimated outflows in PPNe, and compare these with the results of theoretical models. Using Eqn.\,6 of BL14, 
%the accretion rate, \mdotaccr$\propto P/t_{acc}$, and $t_{accr}\sim t_{exp}$ (BL14), 
we scale the accretion rate for IRAS\,16342's HVO relative to that of CRL\,618, and find that, \mdotaccr(IRAS\,16342)$\sim$\,$0.2\times$\mdotaccr(CRL\,618). 
Hence, for IRAS\,16342 (as for 15/19 objects in BL14's sample), Bondi-Hoyle-Lyttleton (BHL) wind 
accretion and wind Roche lobe overflow (M-WRLOF) accretion are ruled out, for 
both white-dwarf and main-sequence accretors. Enhanced RLOF from the primary (i.e., at the Red Rectangle level: BL14), or accretion within CE evolution appear 
to be the most likely candidates for powering the HVO in IRAS\,16342.\\

\noindent {\it Acknowledgements} We are grateful to the late Patrick Huggins, who helped in defining the proposal that led to this study.
We thank an anonymous referee for his/her helpful comments. RS's contribution to this research was carried out at JPL, 
California Institute of Technology, under a contract with NASA. W.V., C.S.C, and G.Q-L. acknowledge support from ERC 
consolidator grant 614264, Spanish MINECO grant AYA2012-32032, and ERC Grant Agreement 610256 (NANOCOSMOS), respectively. This paper uses ALMA dataset  
ADS/JAO.ALMA\#2012.1.00678.S. ALMA is a partnership of ESO (representing its member states), 
NSF (USA) and NINS (Japan), together with NRC (Canada), NSC and ASIAA (Taiwan), and KASI 
(Republic of Korea), in cooperation with the Republic of Chile. 
The Joint ALMA Observatory is operated by ESO, AUI/NRAO and NAOJ.

% 12CO peak of central region at 45.65 km/s; 13CO peak between 44 and 48 km/s, but closer to 44 km/s
% imagename  = 'spw0.fits', freqs correct for 12CO32, dV=2.5 km/s near center vel of CO
% h13cn casa chan 119-122 => 48.93 to 41.31 km/s, imagename = 'spw0/spw0-h13cn43' (dV=2.54 km/s, ch 120: nu=3.45286e+11, 121: nu=3.45289e+11 close to cen vel)

% i16342-spw0-wd21_la.fits
% CTYPE2  = 'FREQ    '
% CRVAL2  =   3.442316268080E+11
% CDELT2  =   2.929666516479E+06

% 12CO32, casa ch 111 (ds9 chan 112)  v=44 km/s, nu=3.45745e+11; casa ch 112 v=48 km/s, nu=3.45741e+11
% h13cn,  
% 12CO32 dashed hor line at v=44 km/s
% i16342-12co-h13cn-pv.ps

% 12co-13co32-h13cn-spec.ps
% elliptical region (pixel): Ellipse[[188, 197], [195, 204]] 0.4 x 0.4 arcsec
% 12co32-off-se-ds9pa223-0p6-spc.ps
% 12co32-off-nw-ds9pa43-0p6-spc.ps
% co32 spectra uses following:
%region (pixel): Ellipse[[171, 180], [178, 187]] => area = pi a b /4 = pi 8*8*(0.05)^2 /4 = 0.1256637 asec^2 
%coordinate: world
%xLabel: radio velocity [km/s]
%yLabel: [Jy] Flux Density
% IRAS_16342-3814_12CO32_image.pbcor.fits-raster
% make 12co32  plots with y-axis= Jy/asec^2, dividing by area= 0.1256637 asec^2

% assuming that the actual outflow velocity is within 5\% of the radial outflow velocity of these blobs

% h13cn-torus.fig uses
%  h13cn43-pa44-pix3.eps (made in casa using PV cuts spw0-h13cn43_PA134_3pix_la, spw0-h13cn43_PA44_3pix_la, cen(casa units) = 183, 193)
% 183, 193 = 16:37:39.935 -38:20:17.10
%  h13cn43-pa134-pix3.eps
%  h13cn43-map-ch128-113.eps (/u4/sah1/sahai/optdata/wfpc2/iras16342/alma/spw0/make-h13cn-chanmap.pro)

% casa viewer max=0.065, min=-0.002 
%Contour levels are :
%  1.2500000E-02  2.5000000E-02  3.7500001E-02  5.0000001E-02  6.2500000E-02
%  7.5000003E-02  8.7499999E-02  0.1000000 
% beam 0.49", 0.27", -78.14 deg
% is $0\farcs49\times0\farcs27$, $PA=-78.1$\arcdeg

% plot uses in gnuplot: "zeroline2.cut" u 1:($2/(3*0.1499)) w l ls 1, zeroline2.cut has 3-sigma level in Jy/beam (3*0.000619 = 1.86e-3)
%"12co32-radial-ch111_c43_c223_w26.cut" u 1:(($2+$4)/(2.0*0.1499)) w l lc 1 lw 2 ,\
%     0.501   0.2502E+00      0.501   0.2612E+00  1.706 Jy/asec^2
%     0.303   0.7248E+00      0.303   0.7367E+00  4.875 Jy/asec^2

%"13co32-radial-ch111_c43_c223_w26.cut" u 1:(($2+$4)/(2.0*0.1713)) w l lc 2 lw 2 ,\
%     0.501   0.2548E+00      0.501   0.2751E+00  1.547 Jy/asec^2
%     0.303   0.6721E+00      0.303   0.6932E+00  3.985 Jy/asec^2

%
%  Jy/bm,bm(asec,asec[-ve:Jy/asec^2]),tau,nu(GHz)
%1.706,-1,-1,10,345.796
%  7.25820608E-13  1.38006252E-15  0.608890235  27.2688217
%1.547,-1,-1,3.33,330.588
%  6.58173873E-13  1.57754349E-15  0.575860977  27.5648022


\begin{thebibliography}{}
\bibitem[Balick \& Frank(2002)]{bf02} Balick, B. \& Frank, A. 2002, ARA\&A, 40, 439
\bibitem[Balick et al.(2013)]{2013ApJ...772...20B} Balick, B., Huarte-Espinosa, M., Frank, A., et al.\ 2013, \apj, 772, 20 
\bibitem[Blackman \& Lucchini(2014)]{2014MNRAS.440L..16B} Blackman, E.~G., \& Lucchini, S.\ 2014, \mnras, 440, L16
\bibitem[Bujarrabal et al.(2001)]{bujetal01} Bujarrabal, V., Castro-Carrizo, A., Alcolea, J., \& S{\' a}nchez Contreras, C.\ 2001, \aap, 377, 868
\bibitem[Castro-Carrizo et al.(2010)]{2010A&A...523A..59C} Castro-Carrizo, A., Quintana-Lacaci, G., Neri, R., et al.\ 2010, \aap, 523, A59 
\bibitem[Cherchneff(2006)]{2006A&A...456.1001C} Cherchneff, I.\ 2006, \aap, 456, 1001 
\bibitem[Claussen et al.(2009)]{2009ApJ...691..219C} Claussen, M.~J., Sahai, R., \& Morris, M.~R.\ 2009, \apj, 691, 219
\bibitem[Cox et al.(2000)]{2000A&A...353L..25C} Cox, P., Lucas, R., Huggins, P.~J., et al.\ 2000, \aap, 353, L25 
\bibitem[Decin et al.(2010)]{2010A&A...516A..69D} Decin, L., De Beck, E., Br{\"u}nken, S., et al.\ 2010, \aap, 516, A69 
\bibitem[Draine \& Hensley(2012)]{2012ApJ...757..103D} Draine, B.~T., \& Hensley, B.\ 2012, \apj, 757, 103
\bibitem[Duari et al.(1999)]{1999A&A...341L..47D} Duari, D., Cherchneff, I., \& Willacy, K.\ 1999, \aap, 341, L47 
\bibitem[Gledhill \& Forde(2012)]{2012MNRAS.421..346G} Gledhill, T.~M., \& Forde, K.~P.\ 2012, \mnras, 421, 346 
\bibitem[Guertler et al.(1996)]{gu96} Guertler, J., Koempe, C., \& Henning, T.\ 1996, \aap, 305, 878
\bibitem[He et al.(2008)]{2008A&A...488L..21H} He, J.~H., Imai, H., Hasegawa, T.~I., Campbell, S.~W., \& Nakashima, J.\ 2008, \aap, 488, L21
\bibitem[Huggins(2007)]{2007ApJ...663..342H} Huggins, P.~J.\ 2007, \apj, 663, 342 
\bibitem[Imai(2007)]{2007IAUS..242..279I} Imai, H.\ 2007, Astrophysical Masers and their Environments, 242, 279 
\bibitem[Imai et al.(2012)]{2012PASJ...64...98I} Imai, H., Chong, S.~N., He, J.-H., et al.\ 2012, \pasj, 64, 98
\bibitem[Lee \& Sahai(2003)]{2003ApJ...586..319L} Lee, C.-F., \& Sahai, R.\ 2003, \apj, 586, 319
\bibitem[Likkel et al.(1992)]{1992A&A...256..581L} Likkel, L., Morris, M., \& Maddalena, R.~J.\ 1992, \aap, 256, 581
\bibitem[Likkel \& Morris(1988)]{1988ApJ...329..914L} Likkel, L., \& Morris, M.\ 1988, \apj, 329, 914 
\bibitem[Mamon et al.(1988)]{1988ApJ...328..797M} Mamon, G.~A., Glassgold, A.~E., \& Huggins, P.~J.\ 1988, \apj, 328, 797
\bibitem[Matt et al.(2006)]{2006ApJ...647L..45M} Matt, S., Frank, A., \& Blackman, E.~G.\ 2006, \apjl, 647, L45 
\bibitem[Murakawa \& Izumiura(2012)]{2012A&A...544A..58M} Murakawa, K., \& Izumiura, H.\ 2012, \aap, 544, A58
\bibitem[Neri et al.(1998)]{1998A&AS..130....1N} Neri, R., Kahane, C., Lucas, R., Bujarrabal, V., \& Loup, C.\ 1998, \aaps, 130, 1
\bibitem[Olofsson et al.(2015)]{2015A&A...576L..15O} Olofsson, H., Vlemmings, W.~H.~T., Maercker, M., et al.\ 2015, \aap, 576, L15 
\bibitem[Sahai et al.(2005)]{2005ApJ...622L..53S} Sahai, R., Le Mignant, D., S{\'a}nchez Contreras, C., Campbell, R.~D., \& Chaffee, F.~H.\ 2005, \apjl, 622, L53 
\bibitem[Sahai et al.(2011)]{smv11} Sahai, R., Morris, M.~R., \& Villar, G.~G.\ 2011, \aj, 141, 134
\bibitem[Sahai et al.(1999b)]{sah_lmzl99} Sahai, R., Te Lintel Hekkert, P., Morris, M., Zijlstra, A., \& Likkel, L.\ 1999, \apjl, 514, L115
\bibitem[Sahai et al.(2007)]{2007ApJ...658..410S} Sahai, R., S{\'a}nchez Contreras, C., Morris, M., \& Claussen, M.\ 2007, \apj, 658, 410
\bibitem[Sahai \& Trauger(1998)]{st98} Sahai, R. \& Trauger, J.T. 1998, AJ, 116, 1357
\bibitem[Sahai et al.(2013)]{2013ApJ...777...92S} Sahai, R., Vlemmings, W.~H.~T., Huggins, P.~J., Nyman, L.-{\AA}., \& Gonidakis, I.\ 2013, \apj, 777, 92
\bibitem[Sahai et al.(2006)]{2006ApJ...653.1241S} Sahai, R., Young, K., Patel, N.~A., S{\'a}nchez Contreras, C., \& Morris, M.\ 2006, \apj, 653, 124
\bibitem[Si{\'o}dmiak et al.(2008)]{2008ApJ...677..382S} Si{\'o}dmiak, N., Meixner, M., Ueta, T., et al.\ 2008, \apj, 677, 382-400
\bibitem[Verhoelst et al.(2009)]{2009A&A...503..837V} Verhoelst, T., Waters, L.~B.~F.~M., Verhoeff, A., et al.\ 2009, \aap, 503, 837

\end{thebibliography}
\end{document}